**Title: No evidence of magma ocean on Io based on Juno/JIRAM data**

**Author List:** Federico Tosi*, Alessandro Mura, Francesca Zambon.

*__Correspondence to:__ Dr. Federico Tosi, E-mail: federico.tosi@inaf.it

**Affiliations:**

*Istituto Nazionale di Astrofisica – Istituto di Astrofisica e Planetologia Spaziali (INAF-IAPS), Via del Fosso del Cavaliere 100, 00133 Rome, Italy.*
Federico Tosi, Alessandro Mura, Francesca Zambon

**Abstract.** A recent paper[1] used infrared images of Io acquired by the Juno/JIRAM instrument to derive a latitudinal dependence of the spectral radiance and conclude that such latitudinal dependence is consistent with a magma ocean model. We challenge their conclusions, and we draw attention to some potential issues with their analysis. In this letter, we will use three arguments to show that: (1) the (ref. [1]) paper uses saturated data; (2) the M-filter of the JIRAM imager is only a weak and incomplete proxy for the total power output; and finally (3) even assuming that the radiance was correctly estimated, the latitudinal dependence of the 4.8-µm spectral radiance is not statistically significant. These facts, taken together, demonstrate that the results presented in (ref. [1]) are not sufficient to confirm consistency with a magma ocean model on Io.

**Saturation**
In the subhead "Saturated data" of the "Methods", (ref. [1]) state: "*JIRAM band radiance values plateau at the point of detector saturation. [...]. We identify and flag all saturated pixels and have created a saturation mask for every frame in every JIRAM observation of Io.*"
Fig. S3 of the Supplementary Information shows an interpolated relation linking the saturation band radiance and the integration time, suggesting that knowledge of the latter is sufficient to identify saturated data in images taken with the M-band filter of the JIRAM optical subsystem. Since in Fig. S3 one of the binding points for the relation: $y = 0.0073 \cdot x^{-1}$ corresponds to an integration time of 1 s of the M-band imager, and knowing that the responsivity of the M-band imager is 2,000,000 DN m$^2$ sr W$^{-1}$ s$^{-1}$, it turns out that (ref. [1]) in their analysis include data with a raw signal up to ($0.0073 \cdot 2 \cdot 10^6 =$) 14,600 Data Number (DN).
In (ref. [2]), the team responsible for the instrument specifies that raw data above 12,000 DN should be used with caution. First, using a saturated data implies not knowing the true radiance but only its lower limit, so saturated data should not be used anyway. Furthermore, a double check on the radiance, overlooked by (ref. [1]), can be performed with the JIRAM spectrometer, whose entrance slit is co-located within the field of view of the M-band filter of the imager subsystem[3]. The spectrometer also covers the 4.54-5.01 µm range of the M-band imager. When no saturation occurs, and once the signal measured by the spectrometer is integrated into the same passband of the imager, the two subsystems measure approximately the same radiance value in the same pixels. However, having a much lower responsivity, the spectrometer saturates at much higher signal levels (see Methods) and can be used to calculate the radiance expected for the imager, yielding Fig. 1.

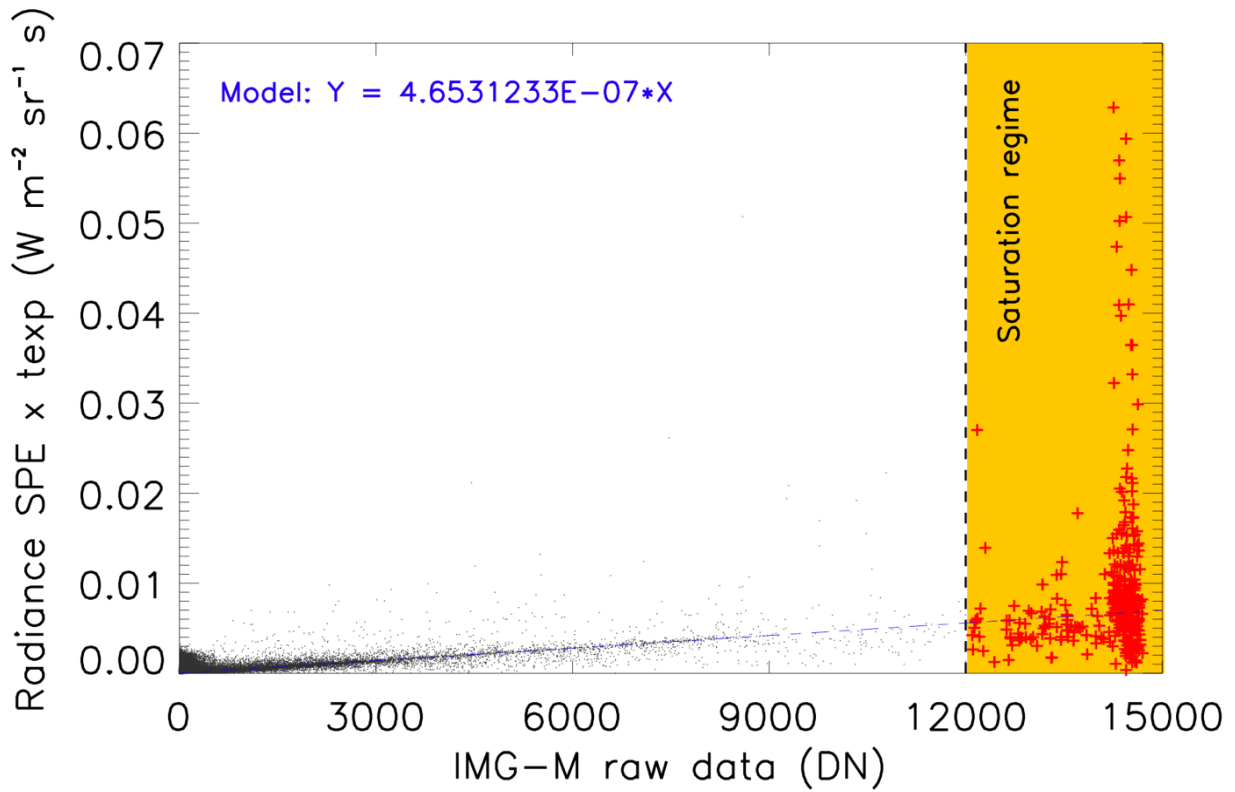

**Fig. 1 Saturation in JIRAM data.** Radiance measured by the JIRAM spectrometer for Io data acquired in orbits 5 to 43, multiplied by the M-band imager integration time, versus Data Numbers (DN) measured in the same pixels by the M-band imager. Points with values >12,000 DN in the imager are highlighted by red crosses on a yellow background. The dashed blue line represents the ideal relationship between the two datasets.

In Fig. 1 one can see that linearity is preserved in the lower part of the dynamics, while above 12,000 DN a regime of saturation or anyway strong detector noise shows up. This proves that using any data above 12,000 is very risky because it can lead to a systematic underestimation of the most intense hot spots. We specify that the data in the figure are real hot spot data used by (ref. [1]); therefore, the high radiance values (red crosses) are data that (ref. [1]) are underestimating up to a factor of ~10 (and in any case are very noisy).

In the caption of Fig. S3 we read: "*we use a threshold of 5% below these values to flag saturated pixels. We also tested a recommended threshold 18% below the shown values (Mura et al., 2020) that flags some additional pixels as saturated, but we found that this makes no significant difference to our results.*"

This procedure does not allow one to reconstruct the true radiance, but only to underestimate it in a different yet similar way. Even admitting that the saturated data were correctly flagged, i.e. using a more conservative threshold of 12,000 DN (as the author correction published on 20 December 2023 and concerning the caption of Fig. S3 would seem to imply), removing the saturated portion and using only the unsaturated portion further underestimates the spectral radiance.

Table 2 of (ref. [1]) reports the radiance values that also include saturated data, but the latter, as already explained, are not significant. This is why (ref. [1]) obtain consistent results with and without applying a saturation mask: both methods lead to a similar underestimation of the radiance of some hot spots, but it is not known by how much.

**Emitted radiance**

In the Introduction, (ref. [1]) state that "*the 4.8-μm spectral radiance is a reasonable proxy for hot spot thermal emission*", suggesting that data returned by the JIRAM M-band filter alone are representative of the total radiance emitted by Io's hot spots.

The following Fig. 2 shows the ratio of the total blackbody radiance to the radiance measured in the JIRAM M-band filter as a function of temperature, for a range of temperature values representative of Io's hot spots. The value of this ratio is not constant and reaches a minimum value of ~13.5 at $T$=769 K. For low temperatures the ratio value increases rapidly, while for high temperatures it can vary between about 15 and 30. Even in the most optimistic case for which there are no temperature values outside those measured in (ref. [2]), the ratio varies from a minimum of ~13.7 at 707 K to a maximum of ~515 at 250 K (this temperature value is probably driven by low spatial resolution, but still indicative of the problem that the temperature is very weakly constrained).

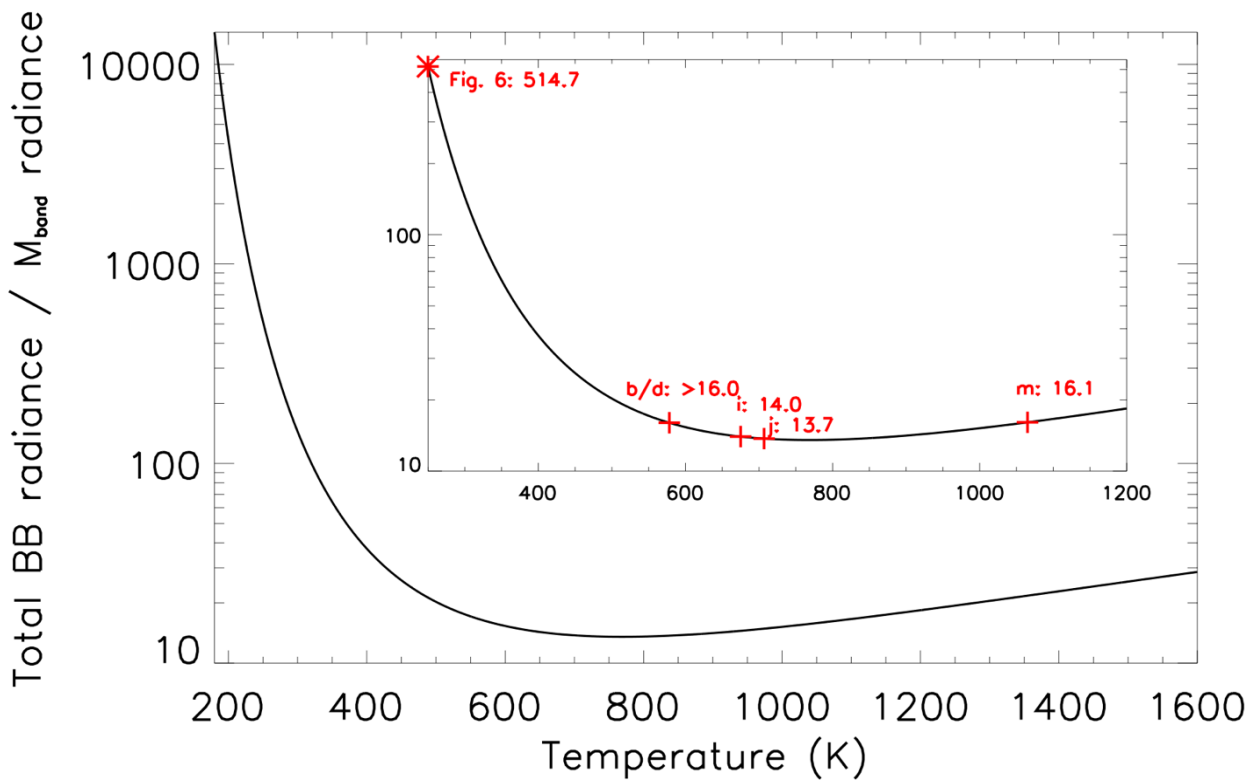

**Fig. 2 Ratio of total blackbody radiance to JIRAM M-band radiance.** Ratio of total blackbody radiance to JIRAM M-band radiance, as a function of Io hot spot temperature (the lower limit at 180 K is dictated by the typical JIRAM in-flight Noise Equivalent Spectra Radiance). The inset is a closeup of the region between 250 and 1200 K, where most data points acquired by JIRAM concentrate. The red crosses on the curve identify the radiance maxima reported in panels *m*, *i*, *j*, *b/d* of Fig. 3 of (ref. [2]), along with the respective values of the $Tot_{BB}/M_{band}$ ratio (in the case of panels *b* and *d* the maximum falls outside the sensitivity range of JIRAM-SPE; therefore, we have very conservatively considered the sensitivity extreme at 5.01 μm even if the corresponding temperature is clearly much lower). The asterisk corresponds to the temperature of 250 K as seen in Fig. 6 of (ref. [2]), which shows Io temperatures, derived with JIRAM-SPE, typical of low spatial resolution data (>100 km/px).

Fig. 2 shows that the assumption that the M-band alone is representative of the total radiance is physically unjustified in (ref. [1]), since Io hot spots encompass a wide range of temperature values. For example, Fig. 3 of (ref. [2]) shows that Io hot spots (same data considered by ref. [1]), analyzed with the JIRAM spectrometer subsystem, display a wide variety of temperature values, with radiance peaks centered at wavelengths from less than 3 µm up to presumably more than 5 µm (i.e. from more than 1000 K to less than 600 K). Furthermore, Fig. 6 of (ref. [2]) highlights other volcanic centers with even lower spectrometer-derived temperatures (this is the case of hot spots observed at very low spatial resolution). (ref. [1]) and (ref. [2]) share some co-authors, suggesting a consensus on those results.

In summary, the "*empirical relationship between 4.8 µm spectral radiance and inferred total hot spot thermal emission*", on which the entire data analysis of (ref. [1]) is based, is contradicted by the physics of the hot spots and by the data measured by JIRAM-SPE. While this assumption has been used in the past where no other binding measurements existed, the JIRAM data, taken as a whole (imager and spectrometer), clearly show its limitations, suggesting that the spectrometer should also be used for this kind of analysis. The standard deviation and skewness reported in Tables 1 and 2 of (ref. [1]) are calculated based on the considered dataset, but without taking into account the major source of error due to the variability of the temperature of hot spots.

**Latitudinal distribution of radiance**

In the "Results" of (ref. [1]), it is stated that polar regions are defined as latitudes >=60° because models of interior tidal heating predict an increase in endogenous heat flux above 60°. Under this assumption, Table 1 and Table 2 of (ref. [1]) list the distribution of hot spots divided into macro-regions: "north polar cap" (60°<lat<90°), "south polar cap" (-90°<lat<-60°), and "lower latitudes" (-60°<lat<60°). The authors conclude that the north polar cap emits more spectral radiance than the south one, and that lower latitudes emit more than the polar regions.

First, we note that using different thresholds, different results are obtained (see Methods). For example, using a threshold of 50°, which gives less unbalanced areas to polar and non-polar regions, we get that the South Pole has the highest radiance per unit area, which makes us suspicious. Hence, to establish a latitudinal dependence in a more rigorous and less arbitrary way, we calculate correlation coefficients between arrays of absolute latitude and spectral radiance density (i.e., spectral radiance observed in given angular bins divided by the area of those bins). By calculating the correlation coefficient between absolute latitude (i.e. independent of sign, which is useful to study the equator/pole asymmetry discussed in ref. [1]) and spectral radiance density for angular bins of different sizes, one can conduct a hypothesis test (Table 1). The absolute value of the correlation coefficient is compared to a critical value depending on the degrees of freedom. The null hypothesis is rejected only if the absolute value of the correlation coefficient exceeds the critical value: in other words, correlations are significant only above this threshold.

**Table 1 Hypothesis test on different angular bins.** The following table uses different angular bins to perform a hypothesis test for latitudinal distribution of spectral radiance density. Columns 1 and 2 report the angular size of the bins and the resulting number of bins, respectively. Column 3 reports the absolute latitude of the bin center at which the maximum spectral radiance density value is recorded. Column 4 reports the absolute value of the correlation coefficient between absolute latitude and spectral radiance density. Column 5 reports the degrees of freedom. Columns 6 and 7 report the critical value to reject the null hypothesis with a confidence level of 95% and 90%, respectively (confidence levels less than 90% are not usually considered). Column 8 reports the outcome of the test.

| Bin size (°) | Number of bins | Absolute latitude of maximum radiance density | Correlation\| coefficient | Degrees of freedom | Threshold .05 (95%) | Threshold .10 (90%) | Result |
|---|---|---|---|---|---|---|---|
| 1 | 90 | 89.5 | 0.0746 | 88 | 0.2074 | 0.1745 | Not significant |
| 2 | 45 | 51.0 | 0.0804 | 43 | 0.2940 | 0.2483 | Not significant |
| 3 | 30 | 52.5 | 0.1790 | 28 | 0.3610 | 0.3061 | Not significant |
| 5 | 18 | 52.5 | 0.2955 | 16 | 0.4683 | 0.4000 | Not significant |
| 10 | 9 | 55.0 | 0.3670 | 7 | 0.6664 | 0.5822 | Not significant |
| 15 | 6 | 52.5 | 0.3355 | 4 | 0.8114 | 0.7293 | Not significant |
| 30 | 3 | 45.0 | 0.7686 | 1 | 0.9969 | 0.9877 | Not significant |

No statistically significant correlation between latitude and spectral radiance density is found with angular bins between 1° and 30°. Moreover, depending on the width of the angular bin, the absolute latitude where the maximum value of spectral radiance density is found can vary from 45° to 89.5°, with a recurring value at 52.5°.

The reason for the lack of statistical significance shall be sought in a characteristic of Io that makes it very difficult to study: in the M-band, 50% of the radiance comes from about 6% of the hot spots considered here (about 17 hot spots in total). In practice, very few hot spots dominate the statistics, making the sample extremely poor. It is therefore evident how the data is prone to statistical variations that are difficult to analyze.

**Summary**

The three arguments presented so far indicate that the results presented by (ref. [1]) and based on Juno/JIRAM data are not sufficient to support a magma ocean model on Io. We suggest that using the complete dataset acquired by the JIRAM instrument, rather than data from the M-band imager alone, would have made the analysis more robust and less prone to data saturation. Data analysis aside, there are critical assumptions from a physical and conceptual point of view, which affect the analysis of the latitudinal variability of Io's power output.

# Methods

## Saturation

The entrance slit of the JIRAM spectrometer subsystem is co-located within the field of view of the M-band filter of the imaging subsystem (see the diagram at ref. [3]). However, the responsivity of the spectrometer is much less than that of the imager (between 4.54 and 5.01 µm it is on average ~16,000 DN m$^2$ µm sr W$^{-1}$ s$^{-1}$, to be compared with 2,000,000 DN m$^2$ sr W$^{-1}$ s$^{-1}$ of the imager). The integration time of the spectrometer is always 1 s. The spectrometer's saturation threshold in terms of spectral radiance is hence always 12,000/16,000 = 0.75 W m$^{-2}$ sr$^{-1}$ µm$^{-1}$ and integrating over the M-band range is ~0.36 W m$^{-2}$ sr$^{-1}$. This value is greater than all points in Fig. S3 of (ref. [1]), an effect of the lesser saturation of the spectrometer subsystem.

To construct the plot in Fig. 1, we take the Io data acquired by the spectrometer at the same time as the M-band imager. For each pixel observed by both the M-band imager and the spectrometer, we first calculate the radiance measured by the spectrometer and integrated in the M-band by summing the signal, measured at a constant integration time of 1 s, in the spectral channels from 4.54 to 5.01 µm, and then multiplying by 0.009 µm (9 nm) which is the average spectral sampling step. We then multiply this radiance by the imager integration time, giving us the radiance the imager should have measured.

## Emitted radiance

To construct Fig. 2, for each temperature between 180 K and 1600 K in steps of 1 K we obtain the total blackbody radiance by first calculating the integral of the Planck function for that temperature between the wavelengths 0 and 10$^6$ µm or 1 m (which for our purposes we can assume to be infinity), and the same integral in the wavelength extremes that represent the full width at half maximum (FWHM) of the JIRAM M-band filter (4.54-5.02 µm). Finally, we plot the value of the ratio between total blackbody radiance and M-band radiance.

## Latitudinal distribution of radiance

**Table 2 Effect of different thresholds on the latitudinal distribution of spectral radiance.** The following table shows the effect of applying different latitude thresholds on the calculation of total radiance and radiance density by latitude bands, using the same dataset as (ref. [1]) in terms of both "*4.8 µm spectral radiance (orbits PJ5 to PJ43) using maximum unsaturated values*" and "*4.8-µm spectral radiance (orbits PJ5 to PJ43) including saturated values*". (ref. [1]) use a 60° threshold to discriminate between high and low latitudes. In addition to this, we explore alternative thresholds at 50° and 40°. For simplicity, we consider Io to be spherical in shape. The area of a spherical zone is calculated as: $A = 2\pi R^2(\sin\varphi_1 - \sin\varphi_2)$, where $R$ is the mean radius of Io equal to 1821.6 km, and $\varphi_1$ and $\varphi_2$ are the latitude extremes. The maximum radiance density value is marked in bold.

| Region of Io | Latitude Range (°) | Area (km$^2$) | Total 4.8-µm unsaturated spectral radiance (GW µm$^{-1}$) | 4.8-µm unsaturated spectral radiance density (kW µm$^{-1}$ km$^{-2}$) | Total 4.8-µm spectral radiance (GW µm$^{-1}$) | 4.8-µm spectral radiance density (kW µm$^{-1}$ km$^{-2}$) |
|---|---|---|---|---|---|---|
| North polar cap | 60–90 | 2.79 · 10$^6$ | 40.79 | 14.6 | 58.69 | 21.0 |
| South polar cap | -60 to -90 | 2.79 · 10$^6$ | 19.80 | 7.1 | 44.52 | 15.9 |
| Lower latitudes | -60 to 60 | 3.61 · 10$^7$ | 946.23 | **26.2** | 1388.09 | **38.4** |
| North polar cap | 50–90 | $4.88 \cdot 10^6$ | 71.42 | 14.6 | 128.58 | 26.4 |
| South polar cap | -50 to -90 | $4.88 \cdot 10^6$ | 156.73 | **32.1** | 185.23 | **38.0** |
| Lower latitudes | -50 to 50 | $3.19 \cdot 10^7$ | 778.70 | 24.4 | 1177.45 | 36.9 |
| North polar cap | 40–90 | $7.45 \cdot 10^6$ | 99.98 | 13.4 | 185.42 | 24.9 |
| South polar cap | -40 to -90 | $7.45 \cdot 10^6$ | 228.50 | **30.7** | 393.30 | **52.8** |
| Lower latitudes | -40 to 40 | $2.68 \cdot 10^7$ | 678.34 | 25.3 | 912.56 | 34.0 |

As can be seen, for thresholds set at 50° and 40° instead of 60°, substantially greater values of total radiance and radiance density are found in the south polar cap compared to the north polar cap.

Let us consider an angular bin $x$ (i.e. $n$ latitude values in steps of $x°$) and calculate the spectral radiance density for the spherical area included in $x°$ of absolute latitude (i.e. we take the latitude value between 0° and $x°$ and between 0° and $-x°$, and so on up to 90°). For each bin, we obtain the spectral radiance density by dividing the maximum 4.8-µm unsaturated spectral radiance of (ref. [1]) by the area of the spherical zone covered by the bin (since this is absolute latitude, we consider double this area, i.e. north plus south).

Once we have obtained two arrays of identical size containing respectively latitude values at the center of the bin and spectral radiance density values, we calculate the Pearson correlation coefficient as the covariance of the two arrays divided by the product of their respective standard deviations.

To undertake a hypothesis test, one must consider the absolute value of the correlation coefficient and compare it with a critical value that depends on the number of degrees of freedom and the confidence level (see e.g. ref. [4]). In this case, the number of degrees of freedom is given by the number of bins minus 2. The most common levels of confidence used are: 90%, 95%, and 99%. We conservatively consider both 90% and 95%; however, in neither of these cases does the test pass.

**Data availability**

The JIRAM dataset used for our analysis is publicly available at the Juno Archive at the Planetary Atmospheres Node (https://pds-atmospheres.nmsu.edu/PDS/data/PDS4/juno_jiram_bundle/data_calibrated/).

4. Critical Values for Pearson's Correlation Coefficient, http://commres.net/wiki/_media/correlationtable.pdf

**Author contributions**
F.T. conceptualized the study and prepared the manuscript and figures. A.M. assisted with conceptualization, interpretation, and manuscript preparation. F.Z. assisted with data analysis.

**Competing interests**
The authors declare no competing interests.

**Additional information**
Correspondence and requests for materials should be addressed to F. Tosi